\documentclass[twocolumn,showpacs,preprintnumbers,amsmath,amssymb]{revtex4}
\usepackage{graphicx}
\usepackage{epsfig}
\usepackage{url}
\usepackage{dcolumn}% Align table columns on decimal point
\usepackage{bm}% bold math

\begin{document}

\title{Universal critical behavior in single crystals and films of YBa$_2$Cu$_3$O$_{7-\delta}$}
 %Insert here a short version of the title if it exceeds 70 characters

\author{Hua Xu}
\author{Su Li}
\author{Steven M.~Anlage}
\author{C.~J.~Lobb}
\altaffiliation[Also at ]{Joint Quantum Institute, University of
Maryland, College Park, MD 20742-4111, USA}
\affiliation{Center for
Nanophysics and Advanced Materials, Department of Physics,
University of Maryland, College Park, MD 20742-4111, USA }

\author{M.~C.~Sullivan}
\affiliation{Department of Physics, Ithaca College, Ithaca, NY
14850, USA}
\author{Kouji Segawa}
\author{Yoichi Ando}
\affiliation{Institute of Scientific and Industrial Research, Osaka
University, Ibaraki, Osaka 567-0047, Japan}

%\date{\today}% It is always \today, today,

\begin{abstract}
We have studied the normal-to-superconducting phase transition in
optimally-doped YBa$_2$Cu$_3$O$_{7-\delta}$ in zero external
magnetic field using a variety of different samples and techniques.
Using DC transport measurements, we find that the dynamical critical
exponent $z=1.54\pm0.14$, and the static critical exponent
$\nu=0.66\pm0.10$ for both films (when finite-thickness effects are
included in the data analysis) and single crystals (where
finite-thickness effects are unimportant). We also measured thin
films at different microwave frequencies and at different powers,
which allowed us to systematically probe different length scales to
avoid finite-thickness effects.  DC transport measurements were also
performed on the films used in the microwave experiments to provide
a further consistency check. These microwave and DC measurements
yielded a value of z consistent with the other results,
$z=1.55\pm0.15$. The neglect of finite-thickness, finite-current,
and finite-frequency effects may account for the wide ranges of
values for $\nu$ and $z$ previously reported in the literature.
\end{abstract}


\maketitle

\section{Introduction}

The high critical temperatures, large penetration depths, and short
coherence lengths of high-temperature superconductors make it
possible to measure critical fluctuations in these materials, in
contrast to conventional superconductors \cite{ffh,chris}. In spite
of nearly two decades of work, however, there is no experimental
consensus on the critical exponents of the superconducting phase
transition in zero magnetic field. By performing both DC and
microwave measurements on thin films, and DC measurements on single
crystals, and doing careful analysis of the data that properly
accounts for finite size, current, and frequency effects, we are
able to provide consistent values for the exponents .

There are two fundamental parameters which characterize a
second-order phase transition such as the superconducting to normal
transition \cite{ffh}. The first is the temperature-dependent
correlation length, $\xi(T)$, which close to the transition
temperature $T_c$ varies as
\begin{equation} \label{eq1}
\xi(T) \sim  |T/T_c-1 |^{-\nu},
\end{equation}
where $\nu$ is the static critical exponent. A second parameter is
the relaxation time $\tau (T)$, which close to $T_c$ varies as
\begin{equation} \label{eq2}
\tau \sim \xi ^{z} \sim  |T/T_c-1 |^{-z\nu},
\end{equation}
where $z$ is the dynamic critical exponent.

It is generally accepted that, theoretically, $\nu\approx 0.67$ in a
superconductor in zero magnetic field, since the phase transition
belongs to the three dimensional (3D) XY universality class
\cite{Hohenberg}. The theoretical situation for the dynamical exponent $z$ is
less certain. Fisher, Fisher, and Huse \cite{ffh} argue that the
number of Cooper pairs is not conserved, so that model A dynamics
\cite{Hohenberg}, which give $z=2$, should apply. Other theoretical
considerations yield $z=1.5$ \cite{Nogueira}, similar to model E
dynamics. Lidmar \cite{Lidmar} and Weber \cite{Weber} present Monte
Carlo simulations that suggest $z\approx 1.5$.

The exponent  $\nu$  can be determined experimentally from a number
of static experiments. In zero field, measurements of penetration
depth,\cite{SKamal,SMAnlage} magnetic
susceptibility,\cite{MBSalamon,APomar,RLiang} specific
heat\cite{MBSalamon,Overend} and thermal expansivity\cite{VPasler}
largely agree that the static critical exponent $\nu\simeq0.67$,
and indicate that the phase transition
in zero field belongs to the 3D-XY universality class. (Note,
however, that there are some measurements which yield different
results.\cite{Inderhees,MBSalamon}

In principle DC conductivity measurements, which depend on both the
statics and the dynamics of the order parameter near $T_c$, can
determine both the static critical exponent $\nu$ and dynamical
critical exponent $z$. The exponents $\nu$ and $z$ are expected to
be universal, but values extracted from conductivity measurements
are not consistent. For example, DC conductivity measurements yield
a wide range of values for critical exponents: $\nu = 0.63$ to 1.2
and $z = 1.25$ to $8.3$.\cite{Yeh,Nojima,Mattfinitesize,Jamesbooth,
Moloni,Roberts,Voss-deHaan}

AC measurements can determine both the real and imaginary parts of
the fluctuation conductivity, providing another probe of critical
dynamics.\cite{Jamesbooth,AlanDorsey,Peligrad,ffh} Measurements over
a broad frequency range allow one to probe the dynamical behavior of
the system and directly measure the fluctuation
lifetime.\cite{Jamesbooth} These experiments are difficult and
seldom done, and the available results are inconsistent, with values
of $z$ ranging from 2 to 5.6. Booth \emph{et al.} investigated the
frequency-dependent microwave conductivity of
$YBa_2Cu_3O_{7-\delta}$(YBCO) films above $T_c$ and obtained $z=2.3$
to $3$.\cite{Jamesbooth} Nakielski \emph{et al.} measured the
conductivity of YBCO at low frequency($<$2 GHz) and obtained
$z\approx5.6$.\cite{Nakielski} Osborn \emph{et al.} did a similar
experiment on $Bi_2Sr_2CaCu_2O_{8+\delta}$ and obtained
$z\approx2$.\cite{KDOsborn} For an optimally doped
$La_{2-x}Sr_xCuO_4$(LSCO) film, Kitano \emph{et al.} found that
their data were consistent with the 3D-XY model with diffusive
dynamics, $\nu\approx0.67,z\approx2$ in a certain temperature
range\cite{Kiano}.

Although the critical exponents $\nu$ and $z$ should be universal,
we see that there is at present no consensus in the literature as to their
values for the zero-field transition in the cuprate superconductors.

In this paper we report the results of a variety of complimentary
experiments which yield independent determinations of the critical
exponents.  In section II, we discuss DC transport measurements on
both thin films and thick single crystals.  We present different
ways to infer each exponent from measurements, and show how
finite-thickness effects in the films can confuse interpretation of
the data in the limit of small currents.  We also show that choice
of the proper range of current can avoid the finite-thickness effects in
films, and show how application of a small magnetic field allows the
determination of $\nu$ in both crystals and films.

In section III, we discuss microwave measurements on thin films.
Just as with DC measurements, microwave measurements require a
non-zero current density. How the applied microwave current density
affects the measured response has not been systematically addressed.
Recently Sullivan \emph{et al.} argued that a finite thickness effects at
low current density was the reason for previous inconsistent results
in DC measurements.\cite{Mattfinitesize} The question of whether a
finite-thickness effect influences the AC measurement and the extracted
critical exponents, as in DC measurements, inspired us to study the
power dependence of the microwave fluctuation conductivity. We find
that several length scales play a role in AC conductivity
measurements, and only after their effects are properly accounted
for can the underlying critical dynamics be understood.  As a
further check, after completing the microwave measurements, we
re-patterned the same samples and performed DC measurements.

When finite-size effects are properly accounted for, we find that
\emph{all} of our results are consistent. As discussed in detail
below, we find $\nu$ is $0.66\pm0.10$ and z is $1.55\pm0.15$.

\section{DC Measurements on Single Crystals and Films}

Our crystals are grown by a flux method using Y$_{2}$O$_3$ crucibles
to ensure crystal purity \cite{ando1,ando2}. The films are prepared
by the pulsed-laser deposition technique at $850$ $^{\circ}$C and
$150$ mbar oxygen pressure on SrTiO$_3$
substrates.\cite{doug,Lisuthesis} Transport measurements were
carried out using the standard four-probe method. The currents were
applied along the $ab$ plane for the films and along the $a$-axis
for the crystals. All connections to the sample are made through
double-T low pass filters to reduce noise \cite{matt_noise}. We
measure the samples inside a cryostat covered with a $\mu$-metal
shield so that the residual magnetic field is less than $2\times
10^{-7}$ T. At 96 K, the resistivity of the crystal is around 70
$\mu \Omega$-cm, and the resistivity of the film is around 90 $\mu
\Omega$-cm.

\begin{figure}
\begin{center}
\epsfxsize=3.8in \epsfysize=5in \epsffile{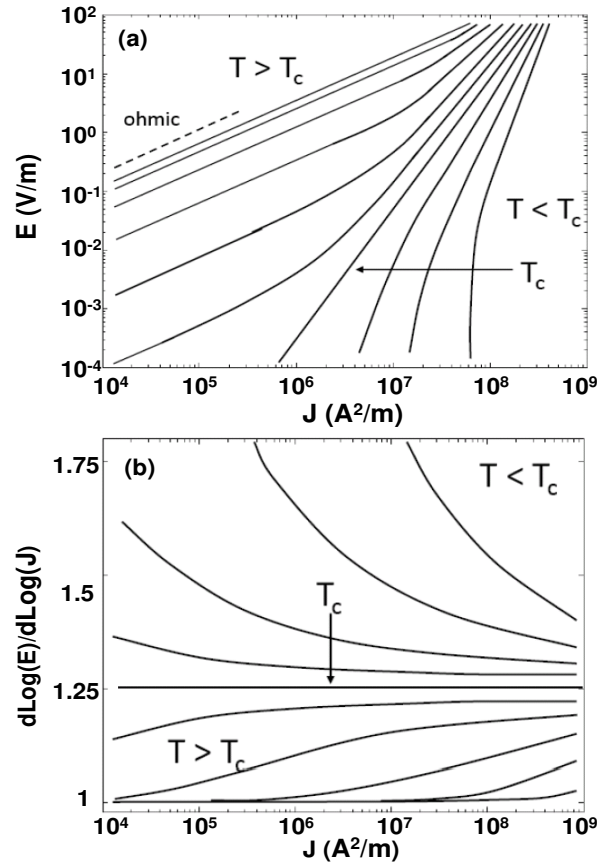}
\end{center}
\caption{Schematic plots of electric field E versus current density
J. (a) E-J plot in log-log scale and (b) $d\log(E)/d\log(J)$ vs. J
in semi-log scale.} \label{EJSchematic}
\end{figure}

In theory, log-log plots of electric field $E$ vs.\ current density
$J$ isotherms above $T_c$ have positive curvatures and display
nonlinearities in the high currents, as shown schematically in Fig.\
\ref{EJSchematic}(a).

As $T_c$ is approached from above, in the limit of $J\rightarrow 0$
\cite{ffh},
\begin{equation}\label{eq:eq25}
\frac{E}{J}\sim \xi^{(D-2-z)}.
\end{equation}
Thus, in a $\log(E)$ vs.\ $\log(J)$ plot, isotherms above $T_c$
exhibit ohmic behavior - a slope of one - at low currents. The
isotherms below $T_c$ have negative curvatures and display vanishing
linear resistance ($R\rightarrow 0$~as~$J\rightarrow0)$. At $T=T_c$,
the critical isotherm is expected to show a power law behavior
\cite{ffh}
\begin{equation}
  E\sim J^{\frac{z+1}{D-1}}
  \label{eq3}
\end{equation}
which is a line with a slope greater than one on a $\log(E)$ vs.\
$\log(J)$ plot.

\begin{figure}
\begin{center}
\epsfxsize=3.4in \epsfysize=4.8in \epsffile{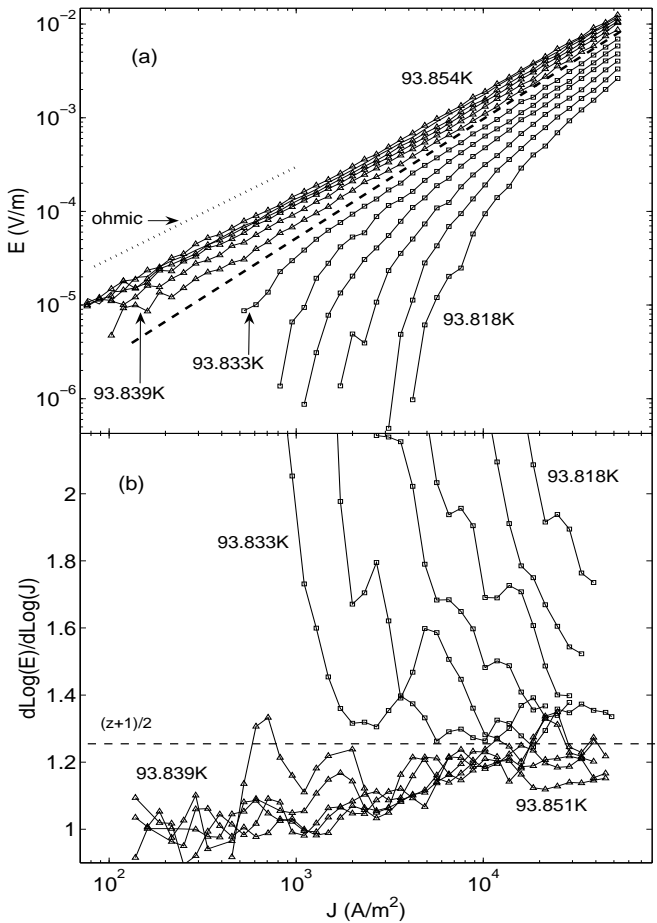}
\end{center}
\caption{(a) The electric field $E$ versus current density $J$ for
an untwinned YBCO single crystal. The dashed line separates the
normal and superconducting phases. The isotherms above the dashed
line exhibits ohmic behavior at low current density. The isotherms
below the dashed line display the vanishing linear resistivity.
Isotherms are separated by 3 mK. (b) Derivative plot of the data in
Fig.\ \ref{EJcrystal}(a). The dashed line separates the normal and
superconducting phases. The crossing of the dashed line with
isotherm of $93.839$ K is due to the noise.  Both plots indicate $z
\approx 1.5$.} \label{EJcrystal}
\end{figure}

In Fig.~\ref{EJcrystal}(a) we show selected $E$ vs.\ $J$ curves in a
log-log plot for an untwinned YBCO single crystal. From the figure,
all isotherms above the dashed line (triangles) have positive
curvatures and have ohmic response in the limit of zero current, and
all isotherms below the dashed line (squares) have negative
curvatures and display vanishing linear resistivity. We can verify
this curvature by fitting the $E-J$ curves to a second-order form of
$\log(E)=a_0 + a_1\log(J)+a_2[\log(J)]^2$, where we use the sign of
$a_2$ to indicate the curvature of each isotherm. We find that $a_2$
is positive above $93.838$ K and negative below $93.836$ K
\cite{note}. Thus, from Fig.~\ref{EJcrystal}(a) we find
$T_c=93.837\pm 0.003$ K. The dashed line in Fig.~\ref{EJcrystal}(a)
separates the superconducting and normal states of the sample and
the high-current part can be fit to $E \sim J^{1.22\pm0.10}$. From
Eq.~(\ref{eq3}) (using $D=3$), we find
\begin{equation}
z = 1.44\pm0.2.
\end{equation}

Another way to evaluate the power law behavior is the
$\big(\frac{\partial \log E}{\partial \log J}\big )_T$ vs.\ $J$ plot
\cite{doug}. From Eq.~(\ref{eq3}), at the transition temperature
$T_c$,
\begin{equation}
\Big (\frac{\partial \log E}{\partial \log J}\Big )_{T_c} =
\frac{z+1}{D-1} \label{eq6}
\end{equation}
thus, the critical isotherm is a horizontal line parallel to the $J$
axis in a derivative plot. The critical isotherm separates the
monotonically increasing isotherms above $T_c$ and the monotonically
decreasing isotherms below $T_c$ in the schematic
Fig.~\ref{EJSchematic}(b). The derivative plot generally displays
the phase transition more clearly than a basic plot of E vs. J.

\begin{figure}
\begin{center}
\epsfxsize=3.4in \epsfysize=2.4in \epsffile{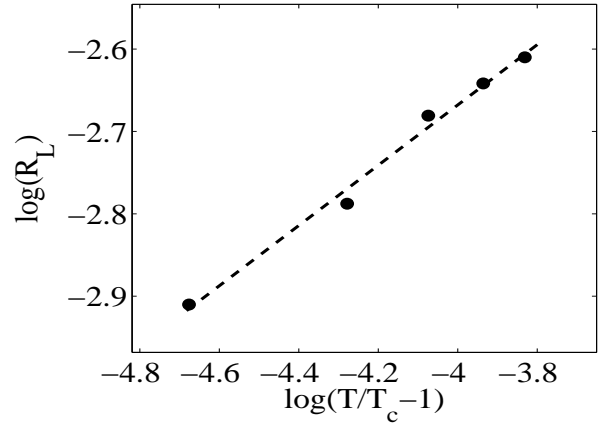}
\end{center}
\caption{Power law fit to Eq.~(\ref{RLequation}) of the ohmic
response of the isotherms just above the transition temperature
$T_c$. Because $R_L \propto (T/T_c-1)^{\nu (z-1)}$, we can use this
line to find $\nu$. By setting $T_c=93.837\pm 0.003$ K and
$z=1.5\pm0.20$, we get $\nu= 0.71\pm0.30$.} \label{rl}
\end{figure}

In Fig.~\ref{EJcrystal}(b), we show a derivative plot of the data
shown in Fig.~\ref{EJcrystal}(a). The dashed line in
Fig.~\ref{EJcrystal}(b) separates the normal and superconducting
phases. The intercept of the dashed line is $1.25\pm 0.10$ and is
expected to be $(z+1)/2$ from Eq.\ (\ref{eq6}). From the derivative
plot, we find
\begin{equation}
z=1.50\pm 0.20,
\end{equation}
which is nearly identical to the result obtained from From
Fig.~\ref{EJcrystal}(a), using Eq.\ (\ref{eq3}).
Fig.~\ref{EJcrystal} (b) also qualitatively shows the change in sign
of $a_2$ discussed above.

\begin{figure}
\begin{center}
\epsfxsize=3.4in \epsfysize=4.8in \epsffile{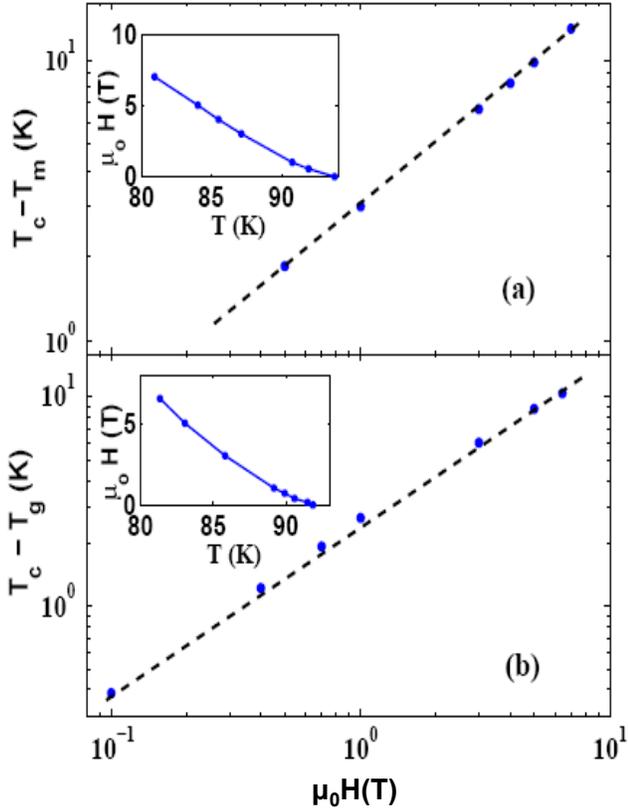}
\end{center}
\caption{(a) $T_c-T_m$ vs.\ $\mu_0H$ of an untwinned YBCO single
crystal. Here, as $T_c-T_{m}(H) \sim H^{1/2\nu}$, we can find $\nu$
from this line without assuming a value for $z$, and find
$\nu=0.68\pm 0.10$. The inset is the melting line for the crystal up
to 7 T. (b) A similar plot of $T_c-T_g$ vs.\ $\mu_0H$ for a YBCO
thin film ($d\approx 150$ nm).  From this curve we find $\nu=0.63\pm
0.10$. The inset is the glass transition line for the film up to 6.5
T.} \label{meltingcrystal}
\end{figure}

The exponent $\nu$ can be found from the low-current ohmic behavior
$R_L$ above $T_c$ by combining Eqs. (\ref{eq1}) and (\ref{eq:eq25})
\cite{ffh}
\begin{equation}
R_L \propto (T/T_c-1)^{\nu (z-1)}. \label{RLequation}
\end{equation}
The slope of the $\log(R_L)$ vs.\ $\log(T/T_c-1)$ plot in
Fig.~\ref{rl}, combined with Eq.\ (\ref{RLequation}), determines
$\nu$. We find
\begin{equation}
\nu=0.71\pm 0.30
\end{equation}
from Fig.\ \ref{rl}.

We can apply a perpendicular magnetic field and look at the
transition in finite field. According to Ref.~\cite{ffh}, the
difference between the critical temperature $T_c$ and the melting
temperature $T_{m(g)}(H)$ is
\begin{equation}
T_c-T_{m(g)}(H) \sim H^{1/2\nu}, \label{eq:melting}
\end{equation}
where $\nu$ is the zero-field static exponent. Eq.\
(\ref{eq:melting}) is expected to be true for clean crystals, where
the transition is a first order melting (m) transition, as well as
for disordered films, where the transition is a glass (g)
transition. We show $T_c-T_m(H)$ vs.\ $\mu_0 H$ in
Fig.~\ref{meltingcrystal}(a) and find
\begin{equation}
\nu=0.68\pm 0.10
\end{equation}
from a power law fit. This result is consistent with the 3D-XY model,
and is also consistent with the result obtained from experiments
using Eq.\ (\ref{RLequation}).

\begin{figure}
\begin{center}
\epsfxsize=3.4in \epsfysize=4.8in \epsffile{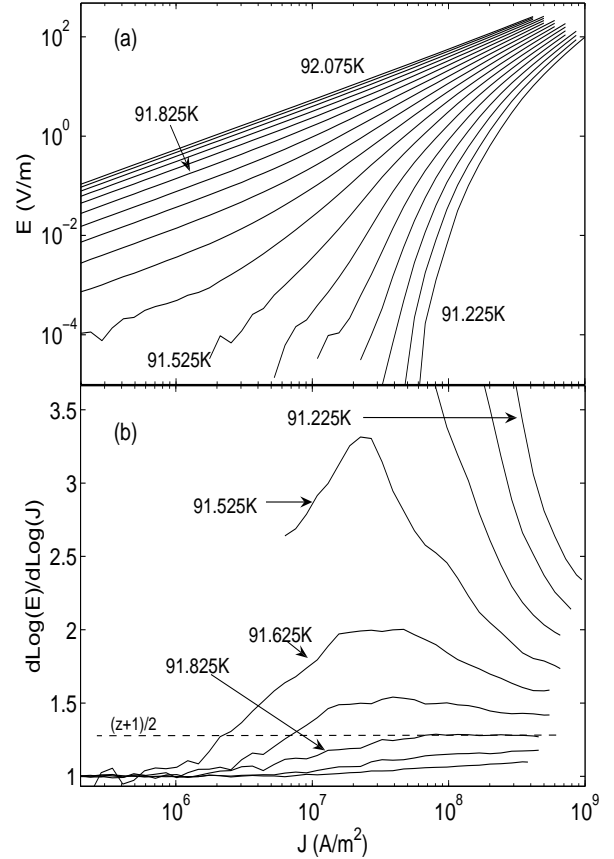}
\end{center}
\caption{(a) $E-J$ isotherms for a 150 nm YBCO film in zero magnetic
field. The spacing between isotherms is 50 mK. (b) The derivative
plot of some selected isotherms from (a). At low current density
regime, the phase transition is obscured by finite-size effects. The
spacing between isotherms is 100 mK.}\label{fig:Ejfilm}
\end{figure}

In Fig.\ \ref{fig:Ejfilm}(a), we show the $E-J$ curves for a $150$
nm thick YBCO $c$-axis oriented optimally-doped film. The isotherms
differ by 0.05 K from 92.075 K to 91.225 K. Unlike
Fig.~\ref{EJcrystal}(a), we cannot find a single straight line in
Fig.~\ref{fig:Ejfilm}(a) that separates the isotherms into two
groups which are either concave or convex exclusively. To help find
the true critical isotherm, we again use the derivative plot
\cite{doug}, where the critical isotherm ideally will correspond to
a horizontal straight line, as in Figs. \ref{EJSchematic}(b) and
\ref{EJcrystal}(b). However, in Fig.~\ref{fig:Ejfilm}(b), there is
no horizontal isotherm, and there are isotherms monotonically
decreasing above $2\times 10^7$ A/m$^2$, and also monotonically
increasing below $2\times 10^7$ A/m$^2$. So, if the experimental
setup were to allow us to measure even smaller voltages, we would
expect all of these isotherms would bend down toward 1 (ohmic behavior) in
the derivative plot at smaller current densities.

The cause of this behavior is most likely finite-size effects
\cite{Mattfinitesize,Woltgens}. Below the transition temperature,
thermal fluctuations take the form of vortex loops.\cite{Nguyenand}
As discussed in the Appendix, vortex loops with length scales of
order
\begin{equation}\label{eq:LJ}
L_J \sim \sqrt{\frac{k_BT}{2\pi\Phi_0J}}
\end{equation}
are probed by current density $J$ \cite{ffh,Woltgens}. The loops
with length scale smaller than $L_J$ will shrink and cause no
dissipation. At high current density, such that $L_J$ is less than
the thickness of the sample $d$, the vortex loops probed in the
experiment are still 3D-like. However, at low current density, such
that $L_J>d$, the size of the vortex loops probed will be limited by
the thickness of the sample and vortex anti-vortex pairs will be
probed. This will lead to a non-diverging energy barrier causing
ohmic behavior even below the bulk transition temperature. When
$L_J$ is equal to 150 nm, the thickness of the film used to produce
the data in Fig.\ \ref{fig:Ejfilm}, the crossover current density is
on the order of $5\times10^6$ A/m$^2$, which is close to where the
isotherms bend towards ohmic behavior in Fig.\ \ref{fig:Ejfilm}(b).

Because of the finite-size effects, the conventional method picks an
incorrect critical isotherm and exponent, contributing to the
inconsistent results from previous transport experiments on
high-T$_c$ films. However, high-current data are not affected by
finite-size effects, and we can extract the $T_c$ and $z$ from the
high-current regime \cite{Mattfinitesize}. If we only look at the
high-current regime in Fig.\ \ref{fig:Ejfilm}(b), it looks very
similar to the schematic derivative plot (Fig.\
\ref{EJSchematic}(b)) and the actual derivative plot of crystal data
(Fig.\ \ref{EJcrystal}(b)). The dashed line in
Fig.~\ref{fig:Ejfilm}(b), which coincides in the high current regime
with the isotherm of $91.825$ K, separates the two phases of the
film. The transition temperature determined from the high-current
regime is $T_c=91.825\pm 0.025$ K, and the intercept of the dashed
line is $1.27\pm 0.07$. According to Eq.\ (\ref{eq6}), from the
high-current data,
\begin{equation}
z=1.54\pm 0.14
\end{equation}
which agrees with the result from the crystal data. In addition, in
our other $c$-axis oriented YBCO films with the thickness $d$
ranging from $100$ nm to $300$ nm, we get consistent values of $z$
ranging from $1.43$ to $1.6$ \cite{note2}. In passing, we note that
one should be cautious about the adverse effect of joule heating
when the measurement is to be done in the high-current regime to
avoid the finite-size effect. In this regard, while the result
reported in Ref. \cite{Ando3} is interesting in that it was the
first I-V measurements of  high-$T_c$ nano-strips, the extracted
critical exponents were likely to be inaccurate because of the
difficulty of avoiding Joule heating in nano-strips at high
currents. In contrast, we have tested heating in our samples by
using the low-frequency technique of Koch \textit{et
al.},\cite{koch} and we have found that heating does not affect the
DC data in samples similar to those measured in this paper at
current densities less than $\approx10^9$ A/m$^2$. \cite{theses}

We are not aware of any way to remove finite-size effects that will
allow us to use Eq.\ (\ref{RLequation}) to determine $\nu$ in films.
Instead, we use Eq.\ (\ref{eq:melting}) which is also applicable to
the vortex-glass transition. By applying a magnetic field we
introduce a magnetic length scale $l_B \propto
\sqrt{\frac{\Phi_0}{B}} $, which is smaller than the film thickness
for $B > 0.1$ T, effectively removing the finite-size limitation in
the film. We show $T_c-T_g(H)$ vs.\ $H$ in
Fig.~\ref{meltingcrystal}(b) and find
\begin{equation}
\nu=0.63\pm 0.10.
\end{equation}

It is important to note that there is more disorder in the film than
in the crystal. Besides the point-like oxygen vacancy disorder as in
the untwinned crystals, there are other kinds of disorder existing
in the film such as twin boundaries and lattice mismatch caused by
the substrate. However, the similar values of $z$ and $\nu$ for the
untwinned crystal and the film argue that the universality of the
phase transition of high-$T_c$ materials is not affected by
disorder.

Hence, taking into account results of DC transport measurements on
both thin films and thick single crystals, we obtained the critical
exponents
\begin{eqnarray}
z &=& 1.54 \pm0.14 \\
\nu &=& 0.66 \pm 0.10.
\end{eqnarray}

\section{Microwave Measurements on Thin Films}

The samples we used for microwave measurements are YBCO films
($d=$100 nm to 300 nm thickness) deposited via pulsed laser
deposition on $NdGaO_3$ and $SrTiO_3$ substrates. AC susceptibility
showed $T_c$ of the films around 90 K with transition widths about
$\Delta T_c=0.2$ K. The resistivity of the films is about $120
\mu\Omega$-cm at 2 K above $T_c$. Using a Corbino reflection
technique, we measured the complex resistivity $\tilde{\rho} =
\rho_1+i\rho_2$ of the samples over a wide frequency range. The
measured complex resistivity is converted to conductivity and the
mean field contribution, as determined from the dc resistivity
measured from room temperature down to the lowest temperature in the
same experiment, is removed.\cite{Jamesbooth,BoothRSI,xuhthesis} The
process is similar to the method described in \cite{Jamesbooth} to
obtain the fluctuation conductivity $\sigma_{fl}$.\cite{ACmeanfield}

\subsection{Frequency Dependent Fluctuation Conductivity and Power Dependence}

According to Fisher-Fisher-Huse (FFH), in zero magnetic field when
the current density is small the complex AC fluctuation conductivity
should scale as \cite{ffh}
\begin{eqnarray}\label{eq:ACscaling}
\sigma_{fl}(T,\omega) & \approx & \xi^{z+2-D}S_{\pm}(\omega \tau).
\end{eqnarray}

In Eq.\ (\ref{eq:ACscaling}), $\xi$ is the correlation length and
$\tau$ is the fluctuation lifetime. The function $S_{\pm}$ is a
universal scaling function above (below) $T_c$ , which should be the
same for all members of a given universality class. As temperature
approaches $T_c$, both $\xi$ and $\tau$ will diverge according to
Eqs.\ (\ref{eq1}) and (\ref{eq2}).

The scaling functions behave as $S_+(y)\rightarrow$ real constant
and $S_{-}(y)\rightarrow 1/(-iy)$ for $y \rightarrow 0$, reflecting
the low frequency behavior above and below $T_c$ respectively. As
$y\rightarrow \infty$, representing $T\rightarrow T_c$,
$S_+(y)\approx S_{-}(y)\approx \tilde{c}y^{(D-2)/(z-1)}$ where
$\tilde{c}$ is a complex constant and D is the dimensionality of the
system. \cite{ffh,AlanDorsey}

\begin{figure}
\epsfxsize=3.4in \epsfysize4.8in \epsffile{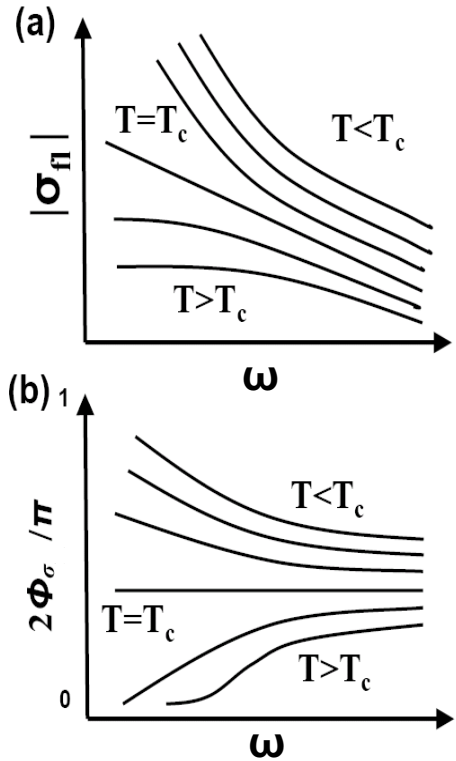} \caption{
Schematic plots of (a) Magnitude $|\sigma_{fl}|$ vs. $\omega$ in
log-log scale and (b) phase $\phi_{\sigma}$ vs. $\omega$ in semi-log
scale at various temperatures around $T_c$, based on
Fisher-Fisher-Huse AC scaling.\cite{ffh}}\label{fig:ACSketch}
\end{figure}

The complex fluctuation conductivity can be written as
$\sigma_{fl}=|\sigma_{fl}|e^{i\phi_{\sigma}}$, so both the magnitude
and phase are predicted to scale
\begin{eqnarray}
|\sigma_{fl}|& \approx & \xi^{z+2-D}|S_{\pm}(\omega \xi^z)|,\\
\phi_{\sigma}& = & \Phi_{\pm}(\omega \xi^z)
\end{eqnarray}
where $\Phi_{\pm}$ is the phase the of the scaling function
$S_{\pm}$. At $T_c$, one expects \cite{AlanDorsey}
\begin{equation}
|\sigma_{fl}|\sim \omega^{-(z+2-D)/z},
\end{equation}
and
\begin{equation}
\phi_{\sigma} = {\pi\over 2}(z+2-D)/z.
\end{equation}
Fig.\ \ref{fig:ACSketch} sketches the expected
Fisher-Fisher-Huse AC scaling behavior of the magnitude and
phase of fluctuation conductivity near $T_c$.\cite{ffh,AlanDorsey}

\begin{figure}
\epsfxsize=3.4in \epsfysize4.8in \epsffile{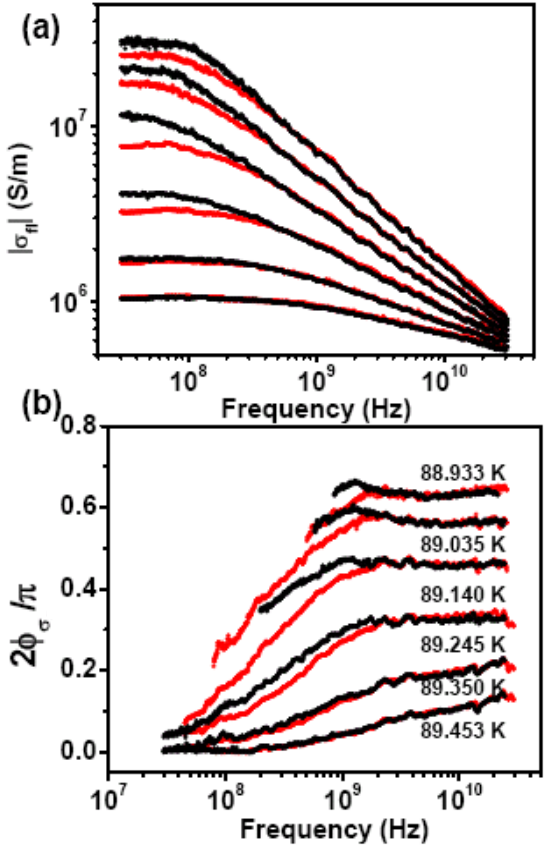} \caption{(Color
online) (a) Magnitude $|\sigma_{fl}|$ and (b) phase $\phi_{\sigma}$
vs. frequency at various temperatures for a typical YBCO
film(xuh139). The black lines were measured with -22dBm power while
the red lines were measured with -2dBm at the same temperature. (For
clarity, only every other isotherm is
shown.)}\label{fig:powercompare}
\end{figure}

Fig.\ \ref{fig:powercompare} shows the measured complex fluctuation
conductivity vs. frequency for various temperatures at two different
microwave powers.  These data display significant and systematic
deviations from the expected FFH scaling sketched in Fig.\
\ref{fig:ACSketch}. At high frequency, both the magnitude and phase
of the fluctuation conductivity look similar to FFH theory. However,
as frequency decreases the measured magnitude of the fluctuation
conductivity below $T_c$ saturates, instead of bending up. All of
the phase isotherms below $T_c$ tend toward zero, indicating ohmic
response, instead of approaching $\pi/2$ at low frequency. These
deviations are qualitatively similar to the low current-density
deviations of E vs. J in DC measurements seen in Fig.\
\ref{fig:Ejfilm}.\cite{Mattfinitesize}

Fig.\ \ref{fig:powercompare} also shows that the applied microwave
power affects the measured fluctuation conductivity, particularly at
low frequencies. As frequency decreases, the higher applied
microwave power decreases the magnitude of the fluctuation
conductivity and depresses the phase. These phenomena cannot be
explained by the AC scaling equation, Eq. (\ref{eq:ACscaling}), and
we need to look at the full version of the FFH dynamic scaling
function, which can be written in the following form with assumed
dimensionality $D=3$ \cite{ffh}:
\begin{eqnarray}
\frac{E}{J} = \xi^{1-z}\chi_{\pm}(J\xi^2,\omega\xi^z,H\xi^2,...).
\label{Eq:FFHGeneral}
\end{eqnarray}
where $E$ is the electric field.

Since the critical point is located in the limit of zero magnetic
field $H$, current density $J$ and frequency $\omega$, increased
applied current should drive the system further away from the
transition and thus into the ohmic regime. In our measurement, the
magnetic field term $H\xi^2$ can be ignored. The two remaining terms
are $J\xi^2$ and $\omega\xi^z$. Qualitatively, at low frequency,
$\omega\xi^z$ is small so that the applied power term, $J\xi^2$, has
more effect on the fluctuation conductivity.

\begin{figure}
\begin{center}
\epsfxsize=3.4in \epsfysize=2.4in \epsffile{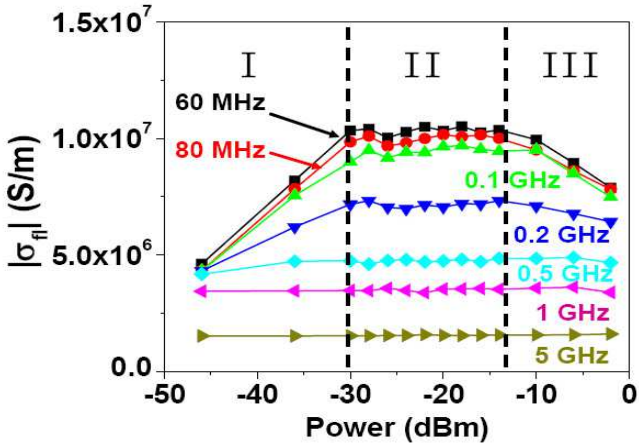}
\end{center}
\caption{(Color online) $|\sigma_{fl}|$ vs. incident microwave power
at different frequencies. ( T=89.140 K, sample xuh139 below
$T_c$)}\label{fig:powerdependence}
\end{figure}

To illustrate the effect of different powers, $|\sigma_{fl}|$ vs.
microwave power at different frequencies is plotted in Fig.\
\ref{fig:powerdependence}. The power dependence of $|\sigma_{fl}|$
clearly varies with frequency. At low frequencies (60 MHz, 80 MHz and
100 MHz), $|\sigma_{fl}|$ vs. incident power increases first as
power increases (Region I) and then saturates (Region II). At very
high power, $|\sigma_{fl}|$ decreases again (Region III). At high
frequencies ($>0.5$ GHz) the fluctuation conductivity is almost power
independent.

The important features in Fig.\ \ref{fig:powerdependence} are that
large applied power affects the fluctuation conductivity, and that
even small power depresses the fluctuation conductivity at low
frequency. While the high-frequency and high-power data in Fig.\
\ref{fig:powerdependence} are consistent with Eq.\
(\ref{Eq:FFHGeneral}), and thus can be explained by FFH scaling
theory\cite{AlanDorsey,Peligrad,ffh}, the low-power low-frequency
behavior is not consistent.

The similarity between this low power and low frequency deviation
and the low current density deviation in DC conductivity measurement
suggests the presence of a "probed length scale" for a finite
frequency. As discussed in connection with Eq.\ (\ref{eq:LJ}) and in
the appendix, when a current with density $J$ is applied, vortex
loops (with large $r$) will "blow out" to infinite size (producing
dissipation). Vortex loops with small $r$ shrink and annihilate (with
no dissipation). A current density induced length scale $L_J$, given
in Eq.\ (\ref{eq:LJ}) separates vortex loops into two categories,
depending on their ultimate fate.

The shrinking of a loop takes time. This time depends on the size of
the loop, thus relating the size of a vortex loop to a time scale.
In AC measurements, small frequency means that large length scales
are probed and \emph{vice versa}. By generalizing the order
parameter relaxation time scale in time-dependent Ginzburg Landau
theory\cite{Tinkham} one can construct a frequency-dependent length
scale
\begin{equation}\label{eq:Lomega}
L_{\omega} = (\frac{ck_BT_c}{\hbar\omega})^{1/z}\xi(0),
\end{equation}
where $c$ is a constant of order 1 and
$\xi(0)/\xi(T)=|T/T_c-1|^{\nu}$.

In AC conductivity measurements, the probed length scale should be
determined by both frequency and current density. Since the smaller
length scale dominates the measured fluctuation conductivity, we
propose a plausible expression for the probed length scale for AC
measurement $L_{AC}$,
\begin{equation}
\frac{1}{L_{AC}} = \frac{1}{L_J}+\frac{1}{L_{\omega}}.
\end{equation}
This formula has the correct limits as $J\rightarrow0$ or
$\omega\rightarrow0$, which corresponds to frequency dependent
$\omega\xi^z$ scaling or current density dependent $J\xi^2$ scaling,
respectively. It is also qualitatively consistent with the two-term
FFH scaling without the magnetic term $H\xi^2$ of Eq.\
(\ref{Eq:FFHGeneral}) in the crossover range. Finite size effects
come into play when $L_{AC}$ approaches the thickness of the film.

\begin{figure}
\begin{center}
\epsfxsize=3.4in \epsfysize=2.4in \epsffile{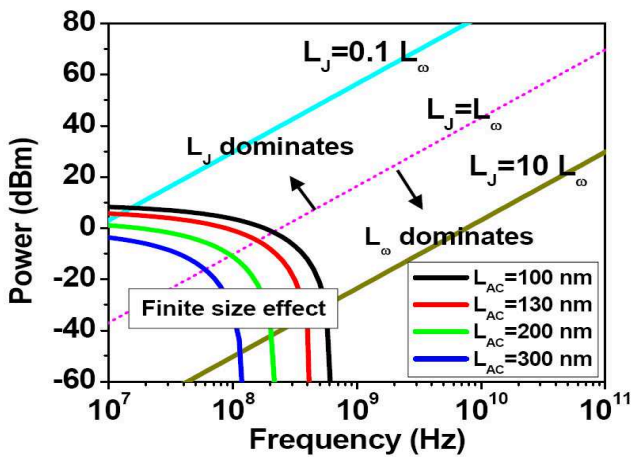}
\end{center}
\caption{(Color online) Summary of length scales and finite size
effects in Corbino AC measurements of fluctuation conductivity of
YBCO films near $T_c$. The dotted line in the figure gives the
boundary $L_J = L_{\omega}$. At low frequency and small current
density, the probed length scale $L_{AC}$ approaches the thickness
of the sample. }\label{fig:AClengthscales}
\end{figure}

Fig.\ \ref{fig:AClengthscales} summarizes the length scales in an AC
measurement in terms of experimental quantities.
\cite{NoteCurrentDensity} In this figure, we use $\xi(0)=5${\AA},
$c=1$ and $z=1.5$. The dotted line in the figure gives the boundary
$L_J = L_{\omega}$. To the right and below the dotted line, when
$L_{\omega}\ll L_J$, the frequency induced length scale dominates,
and one observes mainly frequency dependent scaling of the
fluctuation conductivity. Above the dotted line, when $L_{\omega}\gg
L_J$, current-induced nonlinear effects will dominate the behavior.
This explains the features shown in Fig.\ \ref{fig:powercompare} and
Fig.\ \ref{fig:powerdependence}, where the current density has less
effect on the fluctuation conductivity at high frequency and a
larger effect at low frequency.

At low frequency and small current density, $L_{AC}$ may approach
the thickness of the sample (\emph{d}) or some other length scale
that interrupts the fluctuation vortex loops. Hence deviations from
the simple scaling theory are expected when $L_{AC}>d$.

In our AC measurements, we want to keep to the limit $L_{AC}<d$ to
avoid finite-thickness effect. Hence we choose to stay at low $J$
but high $\omega$. In this region we can find the true critical
behavior without getting into any finite-size effect or crossover
difficulties. Our previous analysis strayed out of this region and
this may account for the larger values of $z$ reported before
\cite{Jamesbooth} and elsewhere in the literature.

\subsection{Improved Data Analysis Method}

In this paper, with very small applied microwave power, -46dBm (corresponding to $J < 2.2\times 10^5$ A/m$^2$), and
high frequency data, we investigated the frequency dependent
fluctuation conductivity around $T_c$. Conventionally, examining
experimental data with the scaling formulas one can search for the
temperature at which the conductivity magnitude best fits to a power
law and has a constant value of $\phi_{\sigma}$, to determine $T_c$
and the dynamic critical exponent $z$. In this analysis process, the
determination of $T_c$ is crucial because it directly affects the
value of $z$. Hence we improved the temperature stability and
conductivity calibration techniques in the experiment, enabling the
measurement of high quality data at small temperature intervals
(50 mK).

Using this data, we improved the conventional data analysis method
\cite{Jamesbooth} to determine $T_c$. One expects a power-law
behavior of $|\sigma_{fl}|$ on frequency at $T_c$, with a change of
curvature on either side (a convex function below $T_c$ and a
concave function above $T_c$), as sketched in Fig.\
\ref{fig:ACSketch}(a). One also expects a plateau in the
conductivity phase vs. frequency at $T_c$, with a change in the sign
of the slope on either side, as sketched Fig.\
\ref{fig:ACSketch}(b).

Unlike the DC I-V curve where Strachan \emph{et al.} used an
opposite concavity criterion to determine $T_c$ in a $dI/dV$
plot,\cite{doug} it is hard to take the frequency derivative of
$|\sigma_{fl}(\omega)|$ because of noise. An alternative approach is
to do a quadratic fit to the data on a log-log plot. Below $T_c$,
the curve bends up with a positive coefficient of $[log(\omega)]^2$
and above $T_c$, the curve bends down with a negative coefficient of
$[log(\omega)]^2$. Hence we did a quadratic fit and found that the
coefficient of the $[log(\omega)]^2$ term changes sign between
temperatures 89.192K and 89.245K, bracketing $T_c$.

The scaling theory also predicts a constant phase angle
$\phi_{\sigma}(\omega)$ at $T_c$. $\phi_{\sigma}(\omega)$ vs. $\log
\omega$ is known to be a decreasing function below $T_c$ and an
increasing function above $T_c$. A linear fit of
$\phi_{\sigma}(\omega)$ vs. $\log f$ also has been done and the
result shows its has negative slope at 89.192K and positive slope at
89.245K, which is consistent with the quadratic fit result of $log
|\sigma_{fl}(\omega)|$ vs. $\log \omega$. The next step is to do a
linear fit for $\log|\sigma_{fl}(\omega)|$ verses $log(\omega)$ to
get the slope of $\log|\sigma_{fl}(\omega)|$, and take the average
of the $\phi_{\sigma}(\omega)$ at $T_c$ to obtain the value of $z$.
From this method, we get the critical temperature $T_c =
89.22\pm0.05 K$ and the critical exponent $z=1.62\pm0.20$.

In addition, we developed a new method to determine $T_c$ from the data. Consider
the Wickham and Dorsey scaling function above $T_c$
\cite{AlanDorsey}
\begin{eqnarray}\label{WDorsey}
S_+(y) &=& \frac{2
z^2[1-\frac{D-2-z}{z}iy-(1-iy)^{(D-2+z)/z}]}{y^2(D-2-z)(D-2)},
\end{eqnarray}
where $ y=\omega \tau\propto\omega \xi^z$. We find at small $y$,
corresponding to temperatures far above $T_c$, the function $S_+(y)$
is essentially independent of dimensionality $D$ and $z$ because the
fluctuation contribution is small. According to Eq.\
(\ref{eq:ACscaling}), one can write
$\sigma_{fl}(T,\omega)=\sigma_0(T)S(\omega/\omega_0)$ where
$\sigma_0(T)$ and $\omega_0(T)$ are characteristic conductivity and
frequency scales, respectively. Both the phase $\phi_{\sigma}(\equiv
\tan^{-1}[\sigma_2^{fl}/\sigma_1^{fl}]$) of $\sigma_{fl}$ and the
magnitude $|\sigma_{fl}|/\sigma_0$ can be treated as scaled
quantities with two temperature-dependent scaling parameters
$\omega_0(T)$ and $\sigma_0(T)$. This is a new data collapse method,
pioneered by Kitano \emph{et al.}\cite{Kiano} They pointed out that
the advantage of this new collapse method is the independence of the
two scaling parameters $\omega_0(T)$ and $\sigma_0(T)$. In this new
data analysis method, the parameters $\omega_0(T)$ and $\sigma_0(T)$
are chosen at each temperature to collapse $\phi_{\sigma}(T)$ vs.
$\omega/\omega_0$ and $|\sigma_{fl}|/\sigma_0(T)$ vs.
$\omega/\omega_0$ to smooth and continuous curves, without \textit{a
priori} determination of $T_c$ or critical exponents.

\begin{figure}[h]
\begin{center}
\epsfxsize=3.6in \epsfysize=2in \epsffile{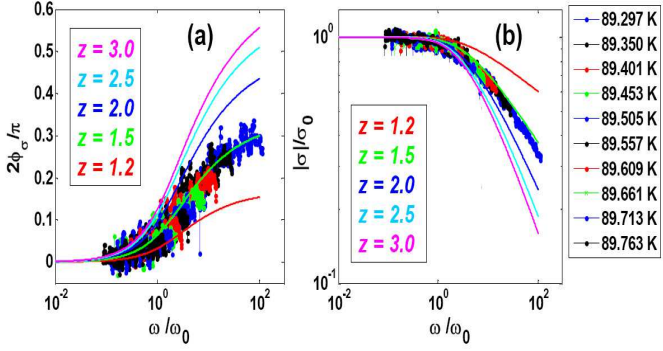}
\end{center}
\caption{Scaling of phase and magnitude of fluctuation conductivity for sample xuh139
to determine $\omega_0(T)$ and $\sigma_0(T)$. (a) $\phi_{\sigma}$
vs. $\omega/\omega_0(T)$; (b) $|\sigma_{fl}|/\sigma_0(T)$ vs.
$\omega/\omega_0(T)$. Solid lines are the theoretical calculation of $|S_+(y))|$ and
$2\phi_{S_+(y)}/\pi$ from Eq.\ (\ref{WDorsey}) for
different values of $z$, assuming $D=3$. These lines can be used to
determine $\omega_0(T)$ and $\sigma_0(T)$ far above $T_c$. In this
figure, only the $\omega_0(T)$ and $\sigma_0(T)$ of the isotherm $T
= 89.763$ K are shown to make the measured
$\phi_{\sigma}(\omega/\omega_0(T))$ and $|\sigma_{fl}|/\sigma_0(T)$
vs. $\omega/\omega_0(T)$ overlap with the theoretical prediction.
For all the other isotherms, $\omega_0(T)$ and $\sigma_0(T)$ for
each temperature are chosen to connect smoothly to the existing
curve of $\phi_{\sigma}(\omega/\omega_0(T))$ and
$|\sigma_{fl}|/\sigma_0(T)$ respectively, to make all the
temperature curves collapse into one smooth and continuous curve.
 }\label{fig:phiomegasig}
\end{figure}

First $\omega_0(T)$ is determined through a collapse plot of
$\phi_{\sigma}$ vs. $\omega/\omega_0(T)$ from high temperature to
low temperature (see Fig.\ \ref{fig:phiomegasig}(a)). Using the
feature that $S_+(y)$ is not sensitive to dimensionality $D$ and $z$ far
above $T_c$, the appropriate $\omega_0(T)$ for isotherms far above
$T_c$ is chosen to make the measured
$\phi_{\sigma}(\omega/\omega_0(T))$ overlap with the theoretical
prediction from the known scaling function $\phi(S_+(y))$. Then at
temperatures closer to $T_c$ where $S_+(y)$ starts to depend on $D$
and $z$, $\omega_0(T)$ for each temperature is chosen to connect
smoothly to the existing curve of
$\phi_{\sigma}(\omega/\omega_0(T))$ and to make all the temperature
curves collapse into one smooth and continuous curve. This process
continues to lower and lower temperature until a temperature is
reached where $\phi_{\sigma}(\omega/\omega_0(T))$ can not be
connected smoothly to the existing curve. In this way, $\omega_0(T)$
for temperature points above $T_c$ can be determined.

To scale the conductivity magnitude, we start with the determined
$\omega_0(T)$ for each temperature, then plot
$|\sigma_{fl}|/\sigma_0(T)$ vs. $\omega/\omega_0(T)$, where
$\sigma_0(T)$, similarly to $\omega_0(T)$, is determined  for each
temperature to make a smooth and continuous curve of
$|\sigma_{fl}|/\sigma_0(T)$ vs. $\omega/\omega_0(T)$.(see Fig.\
\ref{fig:phiomegasig}(b))

\begin{figure}[h]
\begin{center}
\epsfxsize=3.6in \epsfysize=2in \epsffile{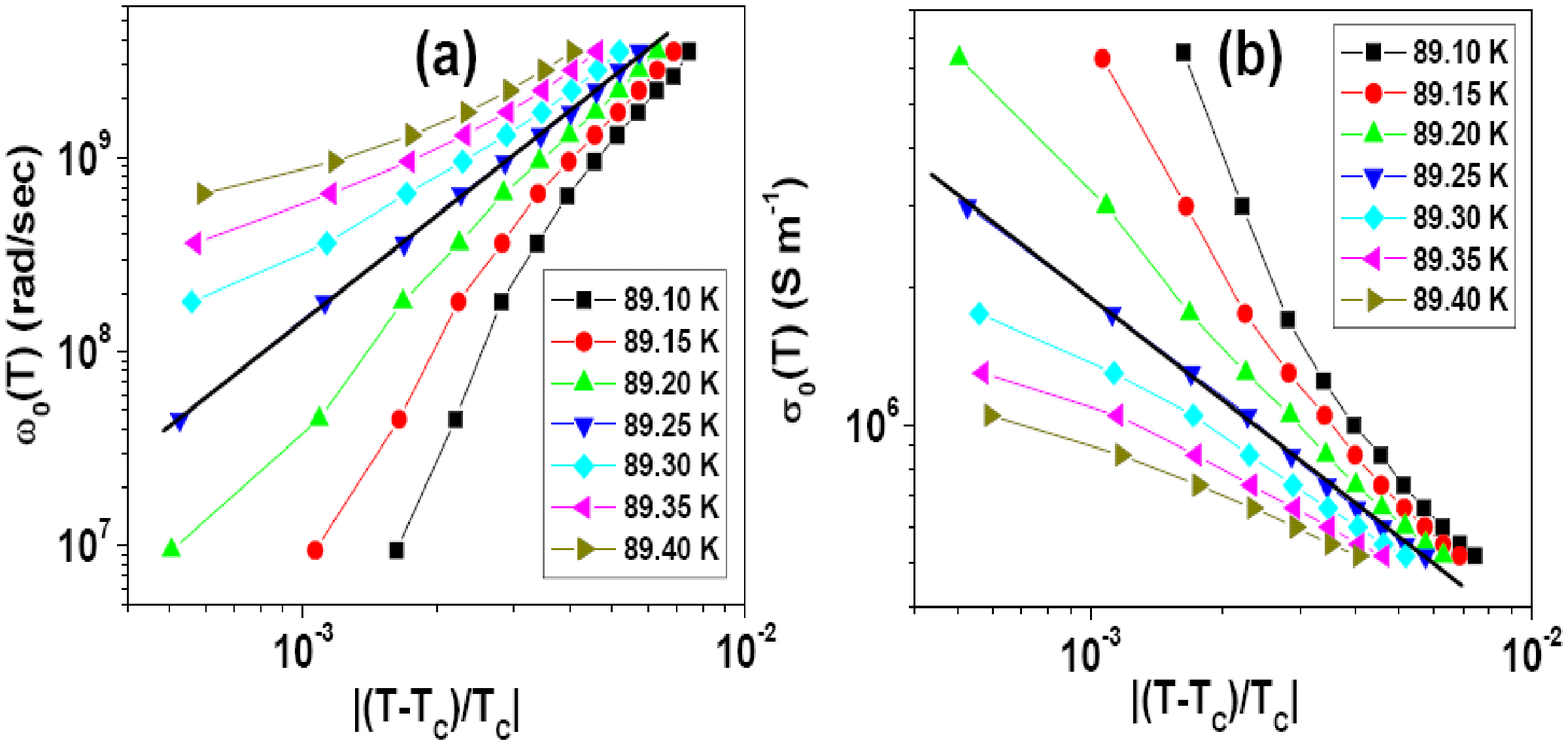}
\end{center}
\caption{$\omega_0(T)$ vs. $|\frac{T-T_c}{T_c}|$ and $\sigma_0(T)$
vs. $|\frac{T-T_c}{T_c}|$ for different assumed $T_c$ for sample
xuh139 and temperature from 89.297K to 89.763K. The errors of
$\omega_0(T)$ and $\sigma_0(T)$ are about the size of the
points.}\label{fig:tomegasig}
\end{figure}

Using the power-law assumption for $\omega_0(T)$ and $\sigma_0(T)$,
$T_c$ can be determined. Fig.\ \ref{fig:tomegasig} shows
$\omega_0(T)$ vs. $t$ and $\sigma_0(T)$ vs. $t$ for different
assumed values of $T_c$. The correct $T_c$ can be determined from
the line showing a pure power-law. Fig.\ \ref{fig:tomegasig}(a)
shows that the blue line which corresponds to an assumed $T_c =
89.25K$ is straightest. Fig.\ \ref{fig:tomegasig}(b) also shows that
the blue line is straightest. From these two figures, $T_c$ is
consistently determined to be $T_c = 89.25\pm0.05K$. This result is
also consistent with the $T_c$ determined by the improved
conventional method.

With the value of $T_c$ determined here, we can do a linear fit for
$\log|\sigma_{fl}(\omega)|$ verses $log(\omega)$ to get the slope of
$\log|\sigma_{fl}(\omega)|$, and take the average of the
$\phi_{\sigma}(\omega)$ at $T_c$ to obtain the value of $z$. Through
this procedure, we obtained the critical exponent $z=1.55\pm0.20$.

In the procedure outlined above, we take advantage of the broad
microwave frequency range of the experiment, which includes
frequencies of order $1/\tau$. High quality data at small
temperature intervals are essential for the implementation of this
method. Another advantage of this method is that many isotherms near
$T_c$ contribute to defining the scaling curve, not just the one
closest to $T_c$.  The new method has the advantage of more
precisely determining $T_c$. So according to the two methods the
critical temperature and exponent for sample xuh139 were determined
to be $T_c = 89.25\pm0.05K,z = 1.55\pm 0.20$.\cite{xuhthesis}

The dynamic critical exponent should be sample independent. To check
the results, we not only repeated measurements on the same sample,
but also repeated the experiment on different samples. Films of
different thickness ($d=$100 nm to 300 nm) were examined, and $z$ was
found to be independent of the thickness, keeping in mind the
constraints of Fig.\ \ref{fig:AClengthscales}. Experiments on 6
samples have been done giving
\begin{equation}
z=1.55\pm0.15.
\end{equation}

\subsection{AC and DC Experiments on the Same Sample}

\begin{figure}
\begin{center}
\epsfxsize=3.4in\epsfysize=4.8in \epsffile{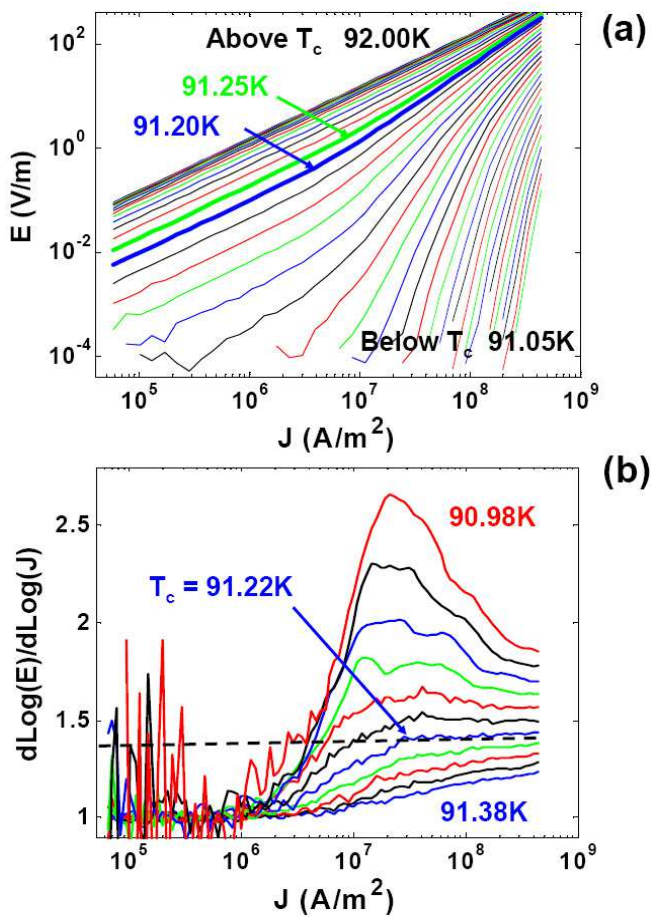}
\end{center}
\caption{(Color online) DC current-voltage characteristics
measurement, performed after the AC experiment on xuh139 in zero
magnetic field. (a) E-J isotherms(50 mK apart), (b) $d\log E/d\log
J$ vs. J derivative plot(40 mK apart).}\label{fig:Figure4}
\end{figure}

We also performed DC current-voltage characteristic measurements on
the same samples.\cite{xuhthesis} Typical results are shown in Fig.\
\ref{fig:Figure4}.(with no background subtraction\cite{Lisuthesis})
According to the negative curvature criterion\cite{doug}, we
determined the critical temperature to be $91.220\pm0.04$ K and the
critical exponent $z=1.75\pm0.2$ from the derivative plot in Fig.\
\ref{fig:Figure4}(b). In Fig.\ \ref{fig:Figure4}, all the isotherms
tend towards ohmic behavior at low current density, brought about by
$L_J>d$ finite-size effects, as discussed for the data shown
previously in Fig.\ \ref{fig:Ejfilm}.\cite{Mattfinitesize} From this
data it is clear that when the current density is smaller than
$1\times10^6 A/m^2$, the sample will have only ohmic response around
$T_c$. The -46dBm applied power in the AC measurement corresponds to
a maximum current density of
$2.2\times10^5A/m^2$($<1\times10^6A/m^2$). This means that for
-46dBm incident power $L_J>d$, verifying a feature of Fig.\
\ref{fig:AClengthscales}, and suggesting that one-parameter scaling
should work when $L_{\omega}<L_J,d$. Hence it is appropriate to
determine $T_c$ and critical exponents with AC data at  -46dBm
applied power.

The difference of $T_c$ between DC and AC measurements is due to the
different thermometer positions and temperature control techniques
of the two experimental systems. The resistance vs. temperature plots from the AC
and DC experiment have a temperature offset about 2.0 K, which is
the difference of the determined $T_c$ from these two methods.

In DC measurements, disorder and heating lead one to systematically
choose a lower temperature isotherm as $T_c$, resulting in an
enhanced value of $z$.\cite{xuhthesis,Lisuthesis} We find that films
with lower normal-state resistivity and smaller $\Delta T_c$ have
smaller values of $z$.\cite{Lisuthesis} The films used in AC
conductivity measurements were grown on $NdGaO_3$ substrates, and
these films have systematically higher resistivity and larger
$\Delta T_c$ than films on $SrTiO_3$ substrates. In addition,
performing DC measurements on the same film after AC measurements
involves more processing steps than a DC measurement alone, and may
result in additional disorder in the sample. This correspondingly
gives larger values of $z$ (Fig.\ \ref{fig:Figure4}). We carefully
repeated the DC measurements alone on different YBCO films grown on
different substrates ($SrTiO_3$ and $NdGaO_3$) and found that the
sample quality does affect the obtained value of
$z$.\cite{xuhthesis} However, for films with high $T_c$, sharp
transition and small resistivity, the obtained value of
$z\approx1.50$, which is consistent with the AC result
$z=1.55\pm0.15$. In addition, DC measurements carried out the same
way on high-quality crystals shown in the previous section also gave
$z\approx1.5$.

\section{Summary and Conclusions}

In this paper, we performed of a variety of complimentary
experiments to determine the critical exponents in optimally-doped
YBa$_2$Cu$_3$O$_{7-\delta}$.

The DC transport measurements on both thin films and thick single
crystals show how finite-thickness effects in the films can confuse
interpretation of the data in the limit of small currents. Only with
the choice of the proper range of current, can one avoid the
finite-thickness effects in films and obtain the correct exponent z,
consistent with the value obtained from thick single crystals
measurements (where finite-thickness effects are unimportant). We
also show how application of a small magnetic field allows the
determination of $\nu$ in both crystals and films. Using DC
transport transport measurements, we find that the dynamical
critical exponent $z=1.54\pm0.14$, and the static critical exponent,
$\nu=0.66\pm0.10$ for both films and single crystals.

Microwave measurements on thin films at different frequencies and at
different powers have also been performed, which allow us to
systematically probe different length scales in the sample. After
developing a comprehensive understanding of length scales in
microwave measurements, we choose to stay at low $J$ but high
$\omega$ to find the true critical behavior without getting into any
finite-size or crossover effects. DC transport measurements were
also performed on the films used in the microwave experiments to
provide a further consistency check. These microwave and DC
measurements yielded a value of $z$ consistent with the other results,
$z=1.55\pm0.15$.

To conclude, using two different measurement methods, we studied the
dynamic fluctuation effects of $YBa_2Cu_3O_{7-\delta}$ single
crystals and thin films around $T_c$. The results of both AC and DC
measurements agree with the XY value for $\nu\approx0.67$ and with
model-E dynamics value for $z=1.55\pm0.15$. \cite{Hohenberg} The
neglect of finite-thickness, finite-power, and finite-frequency
effects may account for the wide ranges of values for $\nu$ and $z$
previously reported in the literature.

\acknowledgments{The authors thank A. T. Dorsey for insightful
discussion. This work has been supported by NSF grant number
DMR-0302596 and by the Maryland Center for Nanophysics and Advanced
Materials. Y. Ando was supported by KAKENHI 19674002 and 20030004,
and K. Segawa by KAKENHI 20740196.}

\newpage

\textbf{\noindent Appendix:  Currents and Length Scales in
Superconductors}
\\
\\
In this appendix we consider a number of length scales in
current-carrying superconductors to provide a clearer physical
meaning for the length scale $L_J$ of Eq.\ (\ref{eq:LJ}).  We first
consider a simple model for fluctuations in superconductors, where
we assume that the only fluctuations are circular vortex loops (or
vortex ``smoke rings'') of radius $r$.\cite{Nguyenand} The energy of
such a loop can be written as

\begin{equation}
   U_{loop} =2\pi r\varepsilon(r)\label{eq:energy}
\end{equation}
where $\varepsilon(r)$ is the energy per unit length of the vortex
loop.  For a straight vortex,
\begin{equation}
 \varepsilon (r=\infty )=\frac{1}{4\pi \mu _{0} } \left(\frac{\Phi _{0} }{\lambda } \right)^{2} K_{0} \left(\frac{\lambda }{\xi } \right)\approx \frac{1}{4\pi \mu _{0} } \left(\frac{\Phi _{0} }{\lambda } \right)^{2} ln\left(\frac{\lambda }{\xi } \right)\label{eq:eps_infty}
\end{equation}
where $K_0$ is a modified Bessel function of the second kind and the
approximate form holds in the limit of high $\kappa \equiv \lambda
/\xi $.\cite{OrlandoAndDelinp}  As a first approximation, we will
assume the energy per unit length is constant, given by Eq.\
(\ref{eq:eps_infty}).

In an infinite superconductor with no applied current, vortex loops
of different sizes occur with different probabilities as thermal
fluctuations.  The probability of finding a loop of size $r$ in a
range $dr$ is given by

\begin{equation}
     P(r)dr=\frac{e^{-\frac{2\pi \varepsilon (r)}{k_{B} T} r} dr}{\int _{\xi }^{\infty }e^{-\frac{2\pi \varepsilon (r)}{k_{B} T} r} dr }
\end{equation}
where interactions between the loops are neglected for simplicity.

We wish to find the size of a typical vortex loop, $r_{med}$. One
way to do this is to find the fraction of loops $f$ with a radius
greater than the median radius, or $r>r_{med}$.  This fraction will
be $f = \frac{1}{2}$,  given by,

\begin{equation}
   f=\frac{\int _{r_{med} }^{\infty }e^{-\frac{2\pi \varepsilon }{k_{B} T} r} dr }{\int _{\xi }^{\infty }e^{-\frac{2\pi \varepsilon }{k_{B} T} r} dr } \equiv \frac{1}{2}\label{eq:f}   .
\end{equation}
 If the energy per unit length of the loop is given by Eq.\ (\ref{eq:eps_infty}), then  Eq.\ (\ref{eq:f}) leads to
\begin{equation}
   r_{med} =\xi +\frac{k_{B} T}{2\pi \varepsilon } \ln2 \label{eq:r_med1} .
\end{equation}
If the second term on the right hand side of Eq.\ (\ref{eq:r_med1})
dominates, this gives

\begin{equation}
   r_{med} \approx \frac{k_{B} T}{2\pi \varepsilon } \ln 2  \Rightarrow \varepsilon \approx \frac{k_{B} T}{2\pi r_{med} } \ln2 \label{eq:r_med2} .
\end{equation}
Eq.\ (\ref{eq:r_med2}) states that, within a factor of $\ln2$, the
total energy of a vortex loop of size $r_{med}$ is equal to $k_BT$,
which is a plausible result.

To check whether the second term on the right side of Eq.\
(\ref{eq:r_med1}) is the dominant one, we combine Eqs.
(\ref{eq:eps_infty}) and (\ref{eq:r_med1}). This leads to
\begin{equation}
 r_{med} =\xi \left[1+\frac{\xi }{\left(\frac{\Phi _{o}^{2} }{4\pi \mu _{0} k_{B} T} \right)} \frac{\ln2}{2\pi } \frac{\kappa ^{2} }{\ln\kappa } \right]=\xi \left[1+\frac{\xi }{\Lambda _{T} } \frac{\ln2}{2\pi } \frac{\kappa ^{2} }{\ln\kappa } \right]\label{eq:long_rmed}
\end{equation}
where  $\Lambda_T$ is defined in Eq.\ (1.1) of Fisher, Fisher, and
Huse \cite{ffh} (in cgs units with the Boltzmann constant $k_B$ defined to be
1). The second terms in Eqs.\ (\ref{eq:r_med1}) and
(\ref{eq:long_rmed}) dominate in the critical regime because $\xi$
diverges while $\Lambda_T$ is fixed.

For simplicity, we drop the $\ln2$ in Eq.\ (\ref{eq:r_med1}), and
use the physically plausible result
\begin{equation}
    r_{med} \equiv \frac{k_{B} T}{2\pi \varepsilon }.\label{eq:r_med3}
\end{equation}

Next consider that a current per unit area $J$ is applied in a
direction perpendicular to the plane of the loop.  The total Lorentz
force acting outward on the loop is
\begin{equation}
   F_{ext} =2\pi rJ\Phi _{0}.
\end{equation}
The energy defined in Eq.\ (\ref{eq:energy}) gives rise to an inward
force that the loop exerts on itself, $-2\pi \varepsilon$. Summing
the forces and finding the point where the  net force is equal to
zero leads to a critical loop size
\begin{equation}
   r_{o} =\frac{\varepsilon }{\Phi _{0} J},\label{eq:ro}
\end{equation}
where, for simplicity, $\varepsilon(r)$ is again assumed to be
independent of $r$.  Note that Eq.\ (\ref{eq:ro}) is \textit{not}
the equation for the current-dependent length scale $L_J$.

Physically, if a vortex loop has $r>r_o$, the external current
``blows out'' the loop to infinite size; this process leads to
dissipation.  If $r<r_o$, the vortex loop shrinks and annihilates.

One can interpret Eq.\ (\ref{eq:ro}) in a different, but equivalent,
way.  The presence of a current density $J$ significantly alters the
population of vortex loops with $r>r_o$, and has less effect on the
vortex loops with $r<r_o$.  In this sense, \textit{a current J
probes the physics on length scales of order $r_o$ and larger.} This
is the type of language that is sometimes used to describe $L_J$.

We next discuss the physical significance of comparing the lengths
$r_o$  and $r_{med}$, Eqs.\ (\ref{eq:r_med3}) and (\ref{eq:ro}).  If
$r_{med}\ll r_o$ , the current is probing a length scale where there
are very few vortex loops.  The current thus acts as a very small
perturbation on the system.  If $r_{med}\gg r_o$, the current is
probing a very short length scale, and a large portion of the
intrinsic vortex population is being disrupted by the current.  The
point where $r_{med}=r_o$ thus marks a crossover in the behavior
from current acting as a small perturbation to current acting as a
large perturbation.

How does a non-infinite film of thickness $d$ affect the physics? It
is plausible to say $r_{med}\ll d$ is the three-dimensional limit,
while $r_{med}\gg d$ is the two-dimensional limit, since in the
second case most of the vortex loops are interrupted by the film
thickness, while in the first case they are not.  This is true as
far as it goes, but misses the key point that an applied current
probes physics at the scale of $r_o$ and larger.  Thus, even in the
limit $r_{med}\gg d$, if $r_o$ is small enough,\textit{ current will
probe physics on length scales smaller than d, and thus the
measurement will not be affected by the finite thickness of the
film.}

In order for the thickness of the film to have a measurable effect,
the current should probe a significant fraction of the loop
population and should also probe lengths on the scale of the film
thickness.  For this to be true, we require

\begin{equation}
  r_{0} =r_{med} \equiv L_{J}\label{eq:equivalence}
\end{equation}
Combining Eqs.\ (\ref{eq:r_med3}),  (\ref{eq:ro}), and
(\ref{eq:equivalence}) gives
\begin{equation}
   L_{J} =\left(\frac{k_{B} T}{2\pi \Phi _{0} J} \right)^{\frac{1}{2} }.\label{eq:Lj}
\end{equation}
This argument leading to Eq.\ (\ref{eq:Lj}) motivates a physical
description for $L_J$:  For any $J$ there is a length scale $L_J$,
given by Eq.\ (\ref{eq:Lj}), such that \textit{roughly half the
equilibrium (zero current) vortex loop  population is strongly
affected by J.}  This is the length that one should compare to the
film thickness for seeing whether or not measurements are in the two
or three dimensional limit.  \textit{The requirements are that there
be a significant fraction of the loops that would exceed the film
thickness, and, in addition, that the current is probing the same
length scale}.

%%%%%%%%%%%%%%%%%%%%%%%%%%%%%%%%%%%%%%

%%%%%%%%%%%%%%%%%%%%%%%%%%%%%%%%%%%%%%


\begin{thebibliography}{33}
\bibitem{ffh}
D. S. Fisher, M. P. A. Fisher, and D. A. Huse, \emph{Phys. Rev. B}
\textbf{43}, 130 (1991); \emph{Nature} \textbf{358}, 553 (1992).

\bibitem{chris}
C. J. Lobb, \emph{Phys. Rev. B} \textbf{36}, 3930 (1987).

\bibitem{Hohenberg}
P. C. Hohenberg and B. I. Halperin, \emph{Rev. Mod. Phys.}
\textbf{49}, 435 (1977).

\bibitem{Nogueira}
F. S. Nogueira and D. Manske, \emph{Phys. Rev. B} \textbf{72},
014541 (2005).

\bibitem{Lidmar}
J. Lidmar, M. Wallin, C. Wengel, S.M. Girvin, and A.P. Young,
\emph{Phys. Rev. B} \textbf{58}, 2827 (1998).

\bibitem{Weber}
H. Weber and H. J. Jensen, \emph{Phys. Rev. Lett.} \textbf{78}, 2620
(1997)

\bibitem{SKamal}
S. Kamal, D.A. Bonn, N. Goldenfeld, P.J. Hirschfeld, R. Liang, and
W.N. Hardy, \emph{Phys. Rev. Lett.} \textbf{73}, 1845 (1994).

\bibitem{SMAnlage}
S.M. Anlage, J. Mao, J.C. Booth, D.H. Wu, and J.L. Peng, \emph{Phys.
Rev. B} \textbf{53}, 2792 (1996).

\bibitem{MBSalamon}
M. B. Salamon, S. E. Inderhees, J. P. Rice, B. G. Pazol, D. M.
Ginsberg and N. Goldenfeld, \emph{Phys. Rev. B} \textbf{38}, 885
(1988); M.B. Salamon, J. Shi, N. Overend, and M.A. Howson,
\emph{Phys. Rev. B} \textbf{47,} 5520 (1993); M.B. Salamon, W. Lee,
K. Ghiron, J. Shi, N. Overend, and M.A. Howson, \emph{Physica A}
\textbf{200,} 365 (1993).

\bibitem{APomar}
A. Pomar, A. Diaz, M.V. Ramallo, C. Torron, and J.A. Veira,
\emph{Physica C} \textbf{218}, 257 (1993).

\bibitem{RLiang}
R. Liang, D.A. Bonn, and W.N. Hardy, \emph{Phys. Rev. Lett.} \textbf{76}, 835
(1996).

\bibitem{Overend}
N. Overend, M.A. Howson, and I.D. Lawrie, \emph{Phys. Rev. Lett.}
\textbf{72}, 3238 (1994).

\bibitem{VPasler}
V. Pasler, P. Schweiss, C. Meingast, B. Obst, H. Wuhl, A.I.
Rykov, and S. Tajima, \emph{Phys. Rev. Lett.} \textbf{81}, 1094
(1998).

\bibitem{Inderhees}
S. E. Inderhees, \textit{et al.}, \emph{Phys. Rev. Lett.}
\textbf{60}, 1178, (1988); S. E. Inderhees \emph{et al.} \emph{Phys.
Rev. Lett} \textbf{66}, 232 (1991).

%\bibitem{Regan}
%S. E. Regan \emph{et al.} \emph{J. Phys. Cond. Matt.} \textbf{3,}
%9245 (1991).

\bibitem{Yeh}
N. C. Yeh, W. Jiang, D. S. Reed, and U. Kriplani, \emph{Phys. Rev.
B} \textbf{47}, 6146 1993.

\bibitem{Roberts}
J. M. Roberts, Brandon Brown, B. A. Hermann, and J. Tate,
\emph{Phys. Rev. B} \textbf{49}, 6890 (1994).

\bibitem{Nojima}
T. Nojima, T. Ishida, and Y. Kuwasawa, Czech. \emph{J. Phys.}
\textbf{46}, Suppl. S3, 1713 (1996).

\bibitem{Jamesbooth}
J. C. Booth, D. Wu, S. B. Qadri, E. F. Skelton, M. S. Osofsky, A.
Pique, and S. M. Anlage, \emph{Phys. Rev. Lett,} \textbf{77}, 4438
(1996).

\bibitem{Mattfinitesize}
M. C. Sullivan, D. R. Strachan, T. Frederiksen, R. A. Ott, M. Lilly,
and C. J. Lobb, \emph{Phys. Rev. B} \textbf{69}, 214524 (2004).

\bibitem{Moloni}
K. Moloni, M. Friesen, S. Li, V. Souw, P. Metcalf, L. Hou, and M.
McElfresh, \emph{Phys. Rev. Lett.} \textbf{78}, 3173 (1997).

\bibitem{Voss-deHaan}
P. Voss-de Haan, G. Jakob and H. Adrian, \emph{Phys. Rev. B}
\textbf{60}, 12443 (1999).

\bibitem{AlanDorsey}
R. A. Wickham and A. T. Dorsey, \emph{Phys. Rev. B} \textbf{61},
6945 (2000).

\bibitem{Peligrad}
D. N. Peligrad, M. Mehring and A. Dul\v{c}i\'{c}, \emph{ Phys. Rev.
B} \textbf{69}, 144516 (2004).

\bibitem{Nakielski}
G. Nakielski, D. G\"{o}rlitz, C. Stodte, M. Welters A. Kr\"{a}mer,
and J. K\"{o}tzler, \emph{Phys. Rev. B} \textbf{55}, 6077 (1997).

\bibitem{KDOsborn}
K. D. Osborn, D. J. Harlingen, V. Aji, N. Goldenfeld, S. Oh, and J.
N. Eckstein , \emph{Phys. Rev. B} \textbf{68}, 144516 (2003).

\bibitem{Kiano}
H. Kitano, T. Ohashi, A. Maeda, and I. Tsukada, \emph{Phys. Rev. B}
\textbf{73}, 092504 (2006).

\bibitem{ando1}
Kouji Segawa and Yoichi Ando, \emph{Phys. Rev. Lett.} \textbf{86},
4907, (2003).

\bibitem{ando2}
Kouji Segawa and Yoichi Ando, \emph{Phys. Rev. B}, \textbf{69}
104521 (2004).

\bibitem{doug}
D. R. Strachan, M. C. Sullivan, P. Fournier, S. P. Pai, T.
Venkatesan, and C. J. Lobb, \emph{Phys. Rev. Lett.} \textbf{87},
067007 (2001).

\bibitem{Lisuthesis}
S. Li, Ph. D. thesis, University of Maryland, 2007
(http://hdl.handle.net/1903/7276).

\bibitem{matt_noise}
M. C. Sullivan, T. Frederiksen, J. M. Repaci, D. R. Strachan, R. A.
Ott, and C. J. Lobb, \emph{Phys. Rev. B} \textbf{70}, 140503 (2004).

\bibitem{note}
The isotherms 93.838 K and 93.836 K are not shown in Fig.
\ref{EJcrystal}(a) for clarity.

%\bibitem{strachan}
%D. R. Strachan \emph{et al.}, Phys. Rev. Lett. \textbf{87}, 067007
%(2001).

\bibitem{Woltgens}
P. J. M. W\"{o}ltgens, C. Dekker, R. H. Koch, B. W. Husse, and A.
Gupta, \emph{Phys. Rev. B} \textbf{52}, 4536 (1995).

\bibitem{Nguyenand}
A. K. Nguyen and A. Sudbo, \emph{Phys. Rev. B} \textbf{60}, 15307
(1999), and references cited there in.

\bibitem{note2}
Our previous work \cite{matt_field} reported $z\approx 2$. This
included a normal-state background subtraction (following W. J.
Skocpol and M. Tinkham, Rep. Prog. Phys. \textbf{38} 1049, (1975)).
This subtraction is not valid in the critical regime, at or near
$T_c$. Our previous work, when reanalyzed, yields $z \approx 1.5$.

\bibitem{Ando3}
Y. Ando, H. Kubota and S. Tanaka, \emph{Phys. Rev. Lett.} \textbf{
69}, 2851 (1992).

\bibitem{koch}
R. H. Koch, V. Foglietti, W. J. Gallagher, G. Koren, A. Gupta, and
M. P. A. Fisher, \emph{Phys. Rev. Lett.} \textbf{63}, 1511 (1989).

\bibitem{theses}
D. R. Strachan, Ph.D. thesis, University of Maryland, 2002, Ch.2; M.
C. Sullivan, Ph.D. thesis, University of Maryland, 2004, Ch. 5.

\bibitem{BoothRSI}
J. C. Booth, D. H. Wu and S. M. Anlage, \emph{Rev. Sci. Instr.}
\textbf{65}, 2082 (1994).

\bibitem{Tinkham}
M. Tinkham, \emph{Introduction to Superconductivity,} McGraw-Hill
Book Co. New York, 1975.

\bibitem{xuhthesis}
H. Xu, Ph. D. thesis, University of Maryland, 2007
(http://hdl.handle.net/1903/7587).

%\bibitem{Aji}
%V. Aji and N. Goldenfeld, \emph{Phys. Rev. Lett.} \textbf{87,}
%197003 (2001).

\bibitem{matt_field}
M. C. Sullivan, D. R. Strachan, T. Fredriksen, R. A. Ott, and C. J.
Lobb, \emph{Phys. Rev. B} \textbf{72}, 092507 (2005).

\bibitem{NoteCurrentDensity}
The sample presents a short circuit termination to the transmission
line, to good approximation, resulting in a relation between applied
microwave power and current density in the Corbino disk, $J(r) =
\sqrt{2P/Z_0}/\pi r t_0$, where $Z_0$ is the characteristic
impedance of the coaxial cable, $t_0$ is the thickness of the
measured sample, and $J(r)$ is the current density amplitude at the
distance $r$ from the center of the Corbino disk. The maximum
current density $J_{max}$ is at the inner radius of the Corbino
disk.

\bibitem{ACmeanfield}
Because the AC mean-field conducitivty remains finite and
well-behaved through $T_c$, and because the critical regime is wide,
a mean field subtraction can be performed with little effect on the
determined critical behavior.

\bibitem{OrlandoAndDelinp}
T. P. Orlando and K. A. Delin, \emph{Foundations of Applied
Superconductivity}, Addison-Wesley, Reading, Massachusetts, 1991, p.
281.


\end{thebibliography}
\end{document}